\documentclass[12pt,headings=big,numbers=noenddot,DIV=14,a4paper]{scrartcl}%

\pdfoutput=1

\usepackage{amsmath,amssymb,amsfonts}
\usepackage[normal,font=small,labelfont=bf,labelsep=period]{caption}
\usepackage[pdftex]{color,graphicx}
\usepackage[english]{babel}
\usepackage[compress]{cite}
\addtokomafont{disposition}{\rmfamily\boldmath}
\usepackage{graphicx}
\usepackage{subfig}
\usepackage{arydshln}

\usepackage[dvipsnames]{xcolor}
\definecolor{darkblue}{rgb}{0,0.2,0.6}
\definecolor{darkgreen}{rgb}{0,0.4,0}

\usepackage[linktoc=page,bookmarks=false,colorlinks=false,
linkbordercolor=RoyalBlue,citebordercolor=ForestGreen,urlbordercolor=CornflowerBlue]{hyperref}

\numberwithin{equation}{section}
\DeclareFontFamily{OT1}{pzc}{}
\DeclareFontShape{OT1}{pzc}{m}{it}{<-> s * [1.10] pzcmi7t}{}
\DeclareMathAlphabet{\mathpzc}{OT1}{pzc}{m}{it}

\title{\vskip 30pt
       \textcolor{black}{Minimal Flavour Violation and SU(5)-unification}}
\author{\normalsize Riccardo
Barbieri and Fabrizio Senia}
\date{\normalsize{\it  Scuola Normale Superiore and INFN, Piazza dei Cavalieri 7, 56126 Pisa, Italy}}

\begin{document}
\begin{titlepage}

\maketitle
\thispagestyle{empty}

\begin{abstract}
\centerline{\bf Abstract}
\noindent
Minimal Flavour Violation in its strong or weak versions, based on $U(3)^3$ and $U(2)^3$ respectively, allows suitable extensions of the Standard Model at the TeV scale to comply with current flavour constraints in the quark sector. Here we discuss considerations analogous to MFV in the context of $SU(5)$-unification, showing the new effects/constraints that arise  both in the quark as in the lepton sector, where quantitative statements can be made controlled by the CKM matrix elements. The case of supersymmetry is examined in detail as a particularly motivated example. Third generation sleptons and neutralinos in the few hundred GeV range are shown to be compatible with current constraints.  
\end{abstract}

\end{titlepage}

\section{Introduction}

The discovery of the Higgs boson and the measurement of some of its couplings, together with the  number of different flavour measurements performed in the last fifteen years or so, have raised the tests of the Standard Model (SM) to a qualitatively higher level. On one side there is the reported evidence for the linear relation, $m_i = \lambda_i v$, between the masses and the couplings to the Higgs boson of the SM particles (for the moment the heavier ones). On the other side several of the flavour-changing SM loops have been experimentally confirmed with strengths as expected in the Cabibbo-Kobayashi-Maskawa (CKM) picture of flavor physics.  Altogether it is appropriate to say that the ensemble of these tests have turned the Yukawa couplings of the Higgs boson in the SM into an element of physical reality. At the same time this strikingly  underlines what is perhaps the major weakness of the SM itself: its inability to predict any of these couplings. This is the current status of the flavour problem in 
the SM, which strongly motivates the efforts to increase the precision  of the mentioned tests, now typically at the $10\div 30\%$ level.

When trying to go beyond the SM, the description of flavour faces a further problem of different nature. If new particles are expected at the TeV scale, the compliance with the flavour tests is highly non trivial. Attempts to achieve it rest on dynamical assumptions, on flavour symmetries or on a  combination thereof. Examples of the first kind are gauge or anomaly mediation of supersymmetry breaking, whereas a combinations of symmetries (typically $U(1)$ factors) and dynamics is invoked in models of alignment. Based on symmetries alone, Minimal Flavour Violation (MFV)  is the way to make new physics at the TeV scale compatible with flavour tests.
 By MFV phenomenologically defined we mean here that in an Effective Field Theory (EFT) approach the only relevant operators are the ones that correspond to the Flavour Changing Neutral Current effects occurring in the SM, weighted  by a common scale and by the standard CKM factors up to (possibly flavour dependent) coefficients  of order unity.  As briefly recalled in the next Section there exist a {\it strong} version of MFV \cite{Chivukula:1987py,Hall:1990ac,D'Ambrosio:2002ex}  based on the $U(3)^3$ flavour group (or equivalent) and a {\it weak} version, based on the $U(2)^3$ flavour group\cite{Barbieri:2011ci,Barbieri:2012uh} (or equivalent).

MFV, as recalled above, refers to the quark sector. Is there something analogous that can be said on the lepton sector, always having in mind new physics not far from the TeV scale? When asking such question,  what comes immediately to mind is the issue of neutrino masses, whose nature (Dirac or Majorana) and origin (at low or high energy among other issues) are unknown\footnote{For some early works extending MFV to the lepton sector see \cite{Cirigliano:2005ck,Davidson:2006bd,Branco:2006hz,Gavela:2009cd}}. This is a difficulty. Perhaps the very small neutrino masses do not influence at all the flavour structure of the charged lepton sector. If so, however,  what is left that can possibly constrain it? In the SM, without neutrino masses, one remains with individual lepton number conservation. With extra particles at the TeV scale individual lepton number conservation is unlikely, but, leaving out neutrino masses, one seems to loose any way to argue further in a truly quantitative way.

In this paper we discuss to what extent  $SU(5)$ unification can avoid this impasse.  By $SU(5)$ unification we simply mean that there exist definite $SU(5)$-invariant Yukawa couplings that give rise, after symmetry breaking, to realistic quark masses and mixings as well as to the observed charged lepton masses. In the low energy theory this leads both to deviations from MFV in the quark sector and to a definite pattern of flavour violation in the charged lepton sector, always controlled by the CKM matrix elements.
The compatibility of such patterns with current bounds will be discussed in general as well as, in particular, considering the possible existence of supersymmetric  particles at the TeV scale.

\section{Strong and weak Minimal Flavour Violation}

Strong MFV \cite{Chivukula:1987py,Hall:1990ac,D'Ambrosio:2002ex} is obtained by assuming the  quark  flavour symmetry  in the gauge sector of the SM 
\begin{equation}
U(3)^{3}\equiv U(3)_{Q}\times U(3)_{u} \times U(3)_{d} 
\end{equation}
to be as well a symmetry of whatever Beyond the Standard Model (BSM) theory under examination,  only broken by the standard Yukawa coupling matrices $Y_u$ and $Y_d$ in the directions
\begin{equation}
Y_{u}=(3,\bar{3},1)\quad\quad Y_{d}=(3,1,\bar{3}).
\end{equation}
 Although this gives up any attempt to understand the pattern of quark masses and mixings, it nevertheless  leads to MFV since, by symmetry transformations,  $Y_u$ and $Y_d$ can be reduced as in the SM to the form
\begin{equation}
Y_u = Y_u^D,\quad\quad Y_d = VY_d^D
\end{equation}
where $Y_{u,d}^D$ are diagonal and $V$ is the CKM matrix. Here with $Y_u$ and $Y_d$ we denote the low energy Yukawa couplings with canonically normalized quark fields, which in general differ from the original symmetry breaking parameters but  have necessarily the same transformation properties under $U(3)^3$ and can equally well be used to characterize the symmetry breaking in the EFT.

Weak MFV \cite{Barbieri:2011ci,Barbieri:2012uh} is based on the observation that 
\begin{equation}
U(2)_{Q}\times U(2)_{u} \times U(2)_{d}\times U(1)_{d3},
\end{equation}
briefly called $U(2)^3$,
is an approximate symmetry of the quark spectrum and mixings\footnote{In fact a larger approximate symmetry of the SM is $U(2)_{Q}\times U(2)_{u} \times U(3)_{d}$\cite{Feldmann:2008ja}. However, implementing MFV with this symmetry leads exactly to the same patterns as $U(3)^3$\cite{Barbieri:2014tja}.}. This suggests to consider $U(2)^3$ rather then $U(3)^3$ as the relevant symmetry with $U(2)^3$ breaking described by small parameters. The only minimal set of spurions that can do this is
\begin{equation}
y_{d3}=(1,1,1)_{-1}\quad\quad \Delta_{u}=(2,\bar{2},1)_{0} \quad\quad \Delta_{d}=(2,1,\bar{2})_{0}\quad\quad \mathbf{V}=(2,1,1)_{0}
\end{equation}
The smallness of these parameters relates to  the smallness of all quark Yukawa couplings except the top one and to the smallness of the elements $V_{td}, V_{ts}, V_{ub}, V_{cb}$  of the CKM matrix. The presence of a doublet, possibly different from $\mathbf{V}$, is necessary to allow the communication between the third and the first two generations. It is simple to convince oneself that any other single doublet transforming differently from $\mathbf{V}$ would have  to be of order unity, thus strongly breaking $U(2)^3$, and would not  lead to MFV.

The low energy Yukawa couplings acquire the form
\begin{equation*}\begin{split}
Y_{u}=\left(
\begin{array}{c:c}
\lambda_{u} & y_{t}x_{t}\mathbf{V} \\ \hdashline
0 & y_{t}.
\end{array}\right)
\quad\quad
Y_{d}=\left(
\begin{array}{c:c}
\lambda_{d} & y_{b}x_{b}\mathbf{V} \\ \hdashline
0 & y_{b}
\end{array}\right)
\end{split}
\end{equation*}
Here $x_{t,b}$ are order one coefficients and, by symmetry transformations, one can set
 \begin{equation}
\mathbf{V} = \begin{pmatrix}0\\ \epsilon\end{pmatrix},\qquad \lambda_u = L_{12}^u\,\lambda_u^D,\qquad
 \lambda_d = \Phi_L L_{12}^d\,\lambda_d^D,
 \label{position}
 \end{equation}
 where $\epsilon$ is a real parameter, $L_{12}^{u,d}$ are rotation matrices in the space of the first two generations with angles $\theta_L^{u,d}$ and $\Phi_L = {\rm diag}\big(e^{-i\phi},1\big)$, i.e. four parameters in total. Due to the smallness of $\epsilon, \lambda_u$ and $\lambda_d$, both $Y_u$ and $Y_d$ are perturbatively diagonalized by pure left transformations, $Y_u= U_u Y_u^D,~ Y_d=U_d Y_d^D$ with $U_{u,d}$ both of the form $U=U_{23}U_{12}$, i.e. the products of two successive unitary transformations in the $1-2$ and $2-3$ sectors.
In turn this gives for the standard CKM matrix
 \begin{equation}\label{CKM}
V_{CKM} = U_u^+ U_d =\begin{pmatrix}
c^u_L c^d_L & \lambda & s^u_L s\,e^{-i\delta}\\
-\lambda & c^u_L c^d_L & c^u_L s\\
-s^d_L s\,e^{i(\delta + \phi)} & -c^d_L s & 1
\end{pmatrix},
\end{equation}
where $s\sim O(\epsilon)$, $s^{u,d}_L=\sin{\theta_L^{u,d}}, c^{u,d}_L=\cos{\theta_L^{u,d}}$ and
$s^u_L c^d_L - s^d_L c^u_L e^{i\phi} = \lambda e^{i\delta}$.
Using this parametrization of the CKM matrix, a direct fit of the tree-level flavour observables, presumably not influenced by new physics,  results in
\begin{align}\label{par}
s^u_L &= 0.096\pm0.008
\,,&
s^d_L &= -0.215\pm0.011
\,,\\
s &= 0.0411 \pm 0.0010
\,,&
\phi &= (95\pm 7)^\circ
\,.
\label{par2}
\end{align}
This  determines the left transformations that diagonalize $Y_u$ and $Y_d$, up to the order one coefficients $x_t, x_b$, and leads to phenomenological MFV. As in the $U(3)^3$ case, the only relevant flavour changing operators are the same as in the SM with coefficients controlled by the CKM factors $\xi_{ij} = V_{ti} V_{tj}^*$. Unlike the $U(3)^3$ case (strong MFV), however, between the $\Delta B =1,2$ and the $\Delta S =1,2$ transitions there are relative $\mathcal{O}(1)$ coefficients, generally complex (weak MFV).

\section{$U(3)^2$ and $SU(5)$-unification}

In terms of the usual trinification of $10 (T) \oplus \bar{5} (\bar{F})$  representations of $SU(5)$, the minimal set of Yukawa couplings for realistic charged fermion masses is
\begin{equation}
\mathcal{L}_Y^{U(3)} = T Y_u T H_5 + T Y_1 \bar{F} H_{\bar{5}} + T Y_2 \bar{F} H_{45} 
\label{superpot}
\end{equation}
where 
$H_5, H_{\bar{5}}, H_{45}$ are Higgs fields transforming under $SU(5)$ as indicated, each with an $SU(2)\times U(1)$ breaking component of similar size, and flavour indices are everywhere left understood. The inclusion of a coupling to $H_{45}$ is necessary to account for the different $\mu - s$ and $e - d$ masses. 
A possibility to describe neutrino masses is to introduce a triplet of fermions, $N$, not transforming under $SU(5)$ and include in $\mathcal{L}_Y $ the further terms $N Y_N \bar{F} H_5 + N M N$.
We assume that the elements of $Y_N$ are small enough not to influence the considerations developed in the following. This is certainly consistent, e.g., if   any of the elements of the matrix $M$ is less than $10^{11}$ GeV. As mentioned, we do this to limit the impact of our ignorance on the values of $Y_N$ and $M$ separately.

In analogy with strong MFV in the SM, the obvious way to go in the $SU(5)$ case is to consider the symmetry
\begin{equation}
U(3)_T\times U(3)_{\bar{F}} \equiv U(3)^2
\end{equation} 
which acts on $T$ and $\bar{F}$ as $T = (3,1)$ and $\bar{F} = (1,3)$\cite{Grinstein:2006cg}.
Furthermore one assumes that $U(3)^2$ is only broken in the directions
\begin{equation}
Y_u = (\bar{6}, 1),\quad\quad Y_1 = (\bar{3}, \bar{3}),\quad Y_2 = (\bar{3}, \bar{3}),
\end{equation}
no matter which other operator is present at whatever scale.

At the TeV scale, after integrating out all heavy particles and including RGE running effects, the Yukawa Lagrangian, written in conventional notation, takes the form
\begin{equation}
\mathcal{L}^{U(3)}_{low~energy}= \bar{Q}_L\lambda_u u_R h + \bar{Q}_L\lambda_d d_R h^* + \bar{L}_L\lambda_e e_R h^* + \dots
\label{lowenergy}
\end{equation}
where $h$ is the only light Higgs doublet and the multiplets $Q, u, d, e$, each with a flavour index and canonically normalized kinetic terms, have definite transformation properties under $U(3)_T\times U(3)_{\bar{F}}$
\begin{equation}
(Q_L, u_R^*, e_R^*)\rightarrow V_T (Q_L, u_R^*. e_R^*),\quad\quad
 (L_L, d_R^*)\rightarrow V_F   (L_L,  d_R^*),
 \end{equation}
as do the low energy Yukawa matrices $\lambda_{u,d,e}$
\begin{equation}
\lambda_u \rightarrow V_T \lambda_u (V_T)^T,\quad (\lambda_d^T, \lambda_e) \rightarrow V_F (\lambda_d^T, \lambda_e) (V_T)^T.
\end{equation}
The matrices $\lambda_{u,d,e}$ control at the same time the flavour symmetry properties of every other higher dimensional operator left understood in (\ref{lowenergy}). This is because they are in the same number as the original "spurions" $Y_u, Y_1, Y_2$, with the same transformation properties up to a complex conjugation. For example it is 
\begin{equation}
\lambda_u = \Sigma_{n >0} ~a_n Y_T^* (Y_T Y_T^*)^n
\end{equation}
where the $a_n$ are complex coefficients of order unity or smaller and we neglect terms involving powers of $Y_1Y_1^+, Y_2 Y_2^+, Y_1 Y_2^+$.

In analogy with the $U(3)^3$ case, by $U(3)^2$ transformations one can set
\begin{equation}
\lambda_u = \lambda_u^D,\quad\quad \lambda_d = V \lambda_d^D,
\end{equation}
where $V$ is again the CKM matrix. On the contrary $\lambda_e$ has the form
\begin{equation}
\lambda_e = V_{eF} \lambda_e^D (V_{eT})^T
\end{equation}
where $V_{eF}, V_{eT}$ are fixed unknown unitary matrices.

\subsection{$U(3)^2$ and Lepton Flavour Violation}

The presence of two spurions $\lambda_d^T$ and $\lambda_e$ with the same transformations properties under $U(3)^2$, one of which dependent on unknown mixing matrices, is the source of potentially large deviations from MFV, particularly from chirality-breaking down-quark operators. If compared with the current bounds, an even stronger direct constraint arises from Lepton Flavour Violation (LFV) and, more specifically, from the $\mu\rightarrow e + \gamma$ transition.

The relevant operator is 
\begin{equation}
\frac{c}{\Lambda^2} e F_{\mu\nu} (\bar{e}_L\lambda_d^T \sigma_{\mu\nu} e_R) v + h.c.
\end{equation}
which, in terms of the physical charged leptons (kept denoted in the same way), becomes
\begin{equation}
\frac{c}{\Lambda^2} e F_{\mu\nu} (\bar{e}_LV_{eF}^+ \lambda_d^D V^T V_{eT}^* \sigma_{\mu\nu} e_R) v + h.c.
\end{equation}
For $\mu\rightarrow e + \gamma$ the likely dominant term is the one proportional to the mass of the b-quark
\begin{equation}
\frac{c}{\Lambda^2} e m_b  F_{\mu\nu} [ A_{32}^* B_{13} (\bar{\mu}_L \sigma_{\mu\nu} e_R)  + 
A_{31}^* B_{23} (\bar{e}_L \sigma_{\mu\nu} \mu_R)] + h.c.; \quad A = V_{eF},\quad B = V_{eT}^+ V,
\end{equation}
which leads to a transition rate
\begin{equation}
\Gamma_{\mu\rightarrow e  \gamma} = \alpha~\frac{m_\mu^3 m_b^2}{\Lambda^4 }
(| c A_{32}^* B_{13}|^2 + | c A_{31}^* B_{23}|^2).
\end{equation}
From the current bound on the Branching Ratio of $5.7\cdot 10^{-13}$\cite{Adam:2013mnn} and using $m_b( 3~\text{TeV})= 2.4~\text{GeV}$, one gets  
\begin{equation}\label{megu3}
|c| \big(|A_{32}^* B_{13}|^2 + |A_{31}^* B_{23}|^2\big)^{1/2} \lesssim 5\cdot 10^{-7} \Big(\frac{\Lambda}{3~\text{TeV}}\Big)^2.
\end{equation}
Even taking into account a possible loop suppression factor, this is a strong constraint, far beyond the typical MFV bounds.
As mentioned a somewhat weaker but still significant constraint on the misalignment of the $\lambda_d$ and $\lambda_e$ matrices arises from the consideration of chirality breaking $\Delta B$ or $\Delta S$  transitions.  Although still highly significant, the constraint in (\ref{megu3}) can be made weaker by a factor $m_b/m_s \approx 50$ if one assumes that all the elements of $Y_2$ in eq. (\ref{superpot})
are at most of order $m_s$, which is sufficient to cure the mass relation problem of the single $T Y_1 \bar{F} H_{\bar{5}}$ coupling.

\section{$U(2)^2$ and $SU(5)$-unification}

The starting point is again the $SU(5)$-invariant Yukawa Lagrangian
\begin{equation}\begin{split}
\mathcal{L}_Y^{U(2)} = y_tT_3 T_3 H_5 + y_t x_t\bold{T}\mathbf{V}T_3 H_5 + \bold{T} \Delta_u \bold{T} H_5  +
 y_b T_3\bar{F}_3 H_{\bar{5}} \\+ y_b x_b \bold{T}\mathbf{V} \bar{F}_3 H_{\bar{5}} 
 + \bold{T}\Delta_1\bar{\bold{F}} H_{\bar{5}} + \bold{T}\Delta_2\bar{\bold{F}} H_{45},
 \end{split}
\label{gutY}
 \end{equation}
invariant under 
\begin{equation}
U(2)_T\times U(2)_{\bar{F}}\times U(1)_{\bar{F}3}\equiv U(2)^2.
\label{U2def}
\end{equation}
With respect to $U(2)^2$ it is
\begin{equation}\label{U2fields}
T_3 = (1,1)_0,\quad \bar{F}_3 = (1,1)_1,\quad \bold{T} = (2,1)_0,\quad \bar{\bold{F}} = (1,2)_0
\end{equation}
with the spurions $\lambda_b, \mathbf{V}, \Delta_{u}, \Delta_{1}, \Delta_{2}$ transforming accordingly to keep $\mathcal{L}_Y$ invariant.

In analogy with the $U(3)^2$ discussion, the low energy Lagrangian in this case assumes the form\footnote{Note a small abuse of notation: here and below the matrices $\lambda_{u,d,e}$ act in the $1-2$ flavour space, unlike the case of Section 3 where they act on the full $1-2-3$ space.}
\begin{equation}\begin{split}
\mathcal{L}^{U(2)}_{low~energy} = (y_t \bar{Q}_{L3} u_{R3} + y_t x_t\bar{\bold{Q}}_L  \mathbf{V}u_{R3}  + y_t x_t\bar{Q}_{L3} \mathbf{V}\bold{ u}_R +
 \bar{\bold{Q}}_L \mathcal{\lambda}_u \bold{ u}_R) h\\
+  (y_b \bar{Q}_{L3} d_{R3} + y_b x_b \bar{\bold{Q}}_L  \mathbf{V} d_{R3} + 
 \bar{\bold{Q}}_L \bold{\lambda}_d \bold{d}) h^*\\
 + ( y_\tau \bar{L}_{L3} e_{R3}+ y_\tau x_\tau \bar{\bold{L}}_L  \mathbf{V} e_{R3} + 
 \bar{\bold{L}}_L \bold{\lambda}_e \bold{e}) h^* + \dots
\end{split}
\label{LEY}
\end{equation}
with self-evident transformation properties under $U(2)^2$ of the various fields/spurions.
By these same transformations one can set $\mathbf{V}, \lambda_u$ and $\lambda_d$ as (see eq. (\ref{position}))
  \begin{equation}
\mathbf{V} = \begin{pmatrix}0\\ \epsilon\end{pmatrix},\qquad \lambda_u = L_{12}^u\,\lambda_u^D (L_{12}^u)^T,\qquad
 \lambda_d = \Phi_L L_{12}^d\,\lambda_d^D,
 \label{position1}
 \end{equation}
 and 
$\lambda_e$ to the form
\begin{equation}
\lambda_e = U_{eF} \lambda_e^D (U_{eT})^T
\end{equation}
where $U_{eF}, U_{eT}$ are fixed unknown $2\times 2$ unitary matrices.  In $3\times 3$ flavour space the low energy Yukawa matrices are given by
\begin{equation}\begin{split}
Y_{u}=\left(
\begin{array}{c:c}
\lambda_{u} & y_{t}x_{t}\mathbf{V} \\ \hdashline
y_{t}x_{t}\mathbf{V}^T & y_{t}.
\end{array}\right)
\quad\quad
Y_{d}=\left(
\begin{array}{c:c}
\lambda_{d} & y_{b}x_{b}\mathbf{V} \\ \hdashline
0 & y_{b}
\end{array}\right)
\quad\quad
Y_{e}=\left(
\begin{array}{c:c}
\lambda_{e} & 0 \\ \hdashline
y_{\tau}x_{\tau}\mathbf{V}^T & y_{\tau}
\end{array}\right)
\end{split}
\label{YuYdYe}
\end{equation}

Altogether this means that to a sufficient approximation the low energy Yukawa matrices are diagonalized by the unitary $3\times 3$ matrices 
\begin{equation}\label{diagU2}
Y_u = U_uY_u^D U_u^T,\quad\quad Y_d = U_d Y_d^D,\quad\quad Y_e = U_{eL}Y_e^D U_{eR}
\end{equation}
where 
\begin{equation}\begin{split}
U_{eL}=\left(
\begin{array}{c:c}
U_{eF}& 0 \\ \hdashline
0& 1
\end{array}\right)
\quad\quad
U_{eR}=\left(
\begin{array}{c:c}
(U_{eT})^T & 0 \\ \hdashline
0 & 1
\end{array}\right) U_{23}(\epsilon)
\end{split}
\label{matru2}
\end{equation}
with $U_{23}(\epsilon)$ a unitary transformation of order $\epsilon$ in the $2-3$ sector. $U_u, U_d$ are the same diagonalization matrices as in normal $U(2)^3$, with their parameters determined as in (\ref{par}, \ref{par2}), except that in $U(2)^3$ case the matrix $U_u$ acts only on the left side  of $Y_u^D$.

Before going further, let us note that $U(2)^2$, unlike $U(3)^2$, makes natural room for the successful  relations $m_b \approx m_\tau$ and $m_\mu \approx 3 m_s$, valid at unification. This only requires that $\Delta_1$ be sufficiently smaller than $\Delta_2$ not to undo the last relation arising from the coupling to $H_{45}$. At the same time $\Delta_1$ must  be capable to give the proper relation between $m_e$ and $m_d$. We assume in the following that all the elements of $\Delta_1$ are at most of the order needed to this purpose.
This in turn implies that  the relative alignment between the $\lambda_d$ and $\lambda_e^T$ matrices is, without any further assumption or tuning, of order $m_d/m_s$ both on the left and on the right side.

\subsection{$U(2)^2$ and LFV}

In analogy with the discussion in the $U(3)^2$ case, the presence of two spurions with the same transformation properties in the down and charged lepton sectors is a source of potentially large flavour violations. In the $U(2)^2$ case, however, there are two significant differences. As just said, in $1-2$ flavour space $\lambda_d$ and $\lambda_e^T$ are misaligned only by relative rotations of order $m_d/m_s$. Furthermore, due to the small $U(2)^2$ breaking,  the diagonalization of both $Y_d$ and $Y_e$ in $2-3$ flavour space is obtained by small rotations of the same order $\epsilon$. 

Here again the leading constraint comes for the $\mu\rightarrow e \gamma$ transition. The relevant operator is 
\begin{equation}
\frac{c^{\mu\rightarrow e \gamma}}{\Lambda^2} e F_{\mu\nu} (\bar{\bold{e}}_L\lambda_d^T \sigma_{\mu\nu} \bold{e}_R) v + h.c.
\label{muegamma}
\end{equation}
only acting in $1-2$ flavour space. Therefore, after going to the physical basis, one obtains a transition rate
\begin{equation}
\Gamma_{\mu\rightarrow e  \gamma} = \alpha~\frac{m_\mu^3 m_s^2}{\Lambda^4 }|c^{\mu\rightarrow e \gamma}|^2
(| A_{22}^* B_{12}|^2 + |A_{21}^* B_{22}|^2),
\end{equation}
where  this time $A, B$ are the misalignment matrices between $\lambda_e$ and $\lambda_D^T$ in the $1-2$ sector, of order $m_d/m_s$.  One gets therefore the bound
\begin{equation}\label{megu2}
|c^{\mu\rightarrow e \gamma}| \Big(\Big| \frac{A_{22}^* B_{12}}{m_d/m_s}\Big|^2 + \Big|\frac{A_{21}^* B_{22}}{m_d/m_s}\Big|^2\Big)^{1/2} \lesssim 5\cdot 10^{-4} \Big(\frac{\Lambda}{3~\text{TeV}}\Big)^2, 
\end{equation}
to be compared with the bound in eq. (\ref{megu3}). This is still a significant limit, but now a loop suppression factor of about $10^{-3}$, as illustrated below in a specific example, could be consistent with new particles of TeV mass.

\subsection{Electric Dipole Moments}

In $U(2)^3$ (as in $U(3)^3$) one expects EDMs for  the quarks, most significantly the ones of the first generation that contribute to the neutron EDM. If one includes also the electron EDM, the relevant operators are\footnote{For brevity we do not discuss the chromo-magnetic  dipole operators for the up and down quarks, but they lead to similar bounds on the corresponding coefficients. The contribution of the charm chromo-electric dipole, $c^{CEDM}_c m_c/\Lambda^2$, to the three gluon Weinberg CP-violating operator gives also a significant bound $|\text{Im}(c^{CEDM}_c)|\lesssim 3\cdot 10^{-2} (\Lambda/\text{3 TeV})^2$\cite{Sala:2013osa}. Note also that in this line by $e, u, d$ we mean specifically the first generation particles and not the flavour   triplets as in Section 3.}
\begin{equation}
\frac{c_e^{EDM} m_e}{\Lambda^2} e F_{\mu\nu} (\bar{e}_L\sigma_{\mu\nu} e_R),\quad 
\frac{c_u^{EDM} m_u}{\Lambda^2} e F_{\mu\nu} (\bar{u}_L\sigma_{\mu\nu} u_R),\quad 
\frac{c_d^{EDM} m_d}{\Lambda^2} e F_{\mu\nu} (\bar{d}_L\sigma_{\mu\nu} d_R)
\label{EDM_U23}
\end{equation}
with $c_{e,u,d}^{EDM}$ generally complex in the physical basis. From the current bounds on the electron and neutron EDMs, respectively $d_e < 8.7\cdot 10^{-29} e\cdot cm$\cite{Baron:2013eja} and $d_n < 2.9\cdot 10^{-26} e\cdot cm$\cite{Baker:2006ts}, one gets the corresponding limits on the imaginary parts of these coefficients\cite{Pospelov:2005pr, Engel:2013lsa} 
\begin{equation}\begin{split}
|\text{Im}(c_d^{EDM})|\lesssim 5.6\cdot 10^{-3} \Big(\frac{\Lambda}{3~\text{TeV}}\Big)^2 \quad\quad
|\text{Im}(c_u^{EDM})|\lesssim 1.6\cdot 10^{-2} \Big(\frac{\Lambda}{3~\text{TeV}}\Big)^2\\ 
|\text{Im}(c_e^{EDM})|\lesssim 8 \cdot 10^{-5} \Big(\frac{\Lambda}{3~\text{TeV}}\Big)^2 \hspace{80pt}
\end{split}
\end{equation}
 In the $U(2)^2$ case, one expects additional contributions to the coefficients of  the operators in line (\ref{EDM_U23}) 
 \begin{equation}
 c_e^{EDM} m_e \rightarrow c_e^{EDM} m_e + \tilde{c}_e^{EDM} m_d,\quad\quad
 c_u^{EDM} m_u \rightarrow c_u^{EDM} m_u + \tilde{c}_u^{EDM} m_t |V_{bu}|^2,
 \end{equation}
 whereas the dipole of the down quark receives negligible corrections.
  For the electron the additional contribution comes from the same type of operator as in eq. (\ref{muegamma}), whereas for the up-quark it is due to the fact that in $U(2)^2$ both $\bold{Q}_L$ and $\bold{u}_R$ transform under the same $U(2)$ group factor or, differently stated, that  the diagonalization matrix $U_u$ is present on both sides of $Y_u^D$ in eq. (\ref{diagU2}).
Barring cancellations this gives the limits
\begin{equation}
|\text{Im}(\tilde{c}_u^{EDM})|\lesssim 1.2\cdot 10^{-2} \Big(\frac{\Lambda}{3\text{TeV}}\Big)^2 \quad\quad
|\text{Im}(\tilde{c}_e^{EDM})|\lesssim 1.6\cdot 10^{-5} \Big(\frac{\Lambda}{3\text{TeV}}\Big)^2\quad. 
\end{equation}

\subsection{$U(2)^2$ and Quark Flavour Violation}

The counterpart in the down-quark sector of the chirality breaking effect in  $\mu\rightarrow e \gamma$ of Sec. 4.1 is due to the operator
\begin{equation}
\frac{c^{\Delta S=1}}{\Lambda^2} g_s G^a_{\mu\nu} (\bar{\bold{d}}_L\lambda_e^T \sigma_{\mu\nu} T^a \bold{d}_R) v + h.c.
\end{equation}
By a similar line of reasoning to the one that leads to the limit in eq. (\ref{megu2}) and following the analyses of \cite{Mertens:2011ts,Barbieri:2012bh}, the effect of this operator on the parameter $\epsilon^\prime$ sets the bound
\begin{equation}
|c^{\Delta S=1}|| \sin{\phi}| \lesssim 6\cdot 10^{-2} \Big(\frac{\Lambda}{3~\text{TeV}}\Big)^2,
\end{equation}
where the factor $\sin{\phi}$ is there to remember the role of phases, in general a combination of them, in this effect and we have set $|A_{12}|\approx |B_{12}|\approx m_d/m_s$.

In the up-quark sector the Yukawa couplings in $3\times 3$ flavour space have the form $Y_u = U_u Y_u^D U_u^T$ with $U_u = U_{12} U_{23}$. Similarly one can write down a $U(2)^2$-invariant dipole operator with the flavour structure $D_u = \tilde{U}_u D_u^D \tilde{U}_u^T$ and $\tilde{U}_u = U_{12} \tilde{U}_{23}$. The point is that  the unitary transformations in the $1-2$ sector are the same in the two cases whereas $U_{23}$ and $\tilde{U}_{23}$, although both of order $\epsilon$, are different from each other. In the physical basis, therefore, $D_u$ keeps to a good approximation the same form, except for a different $U_{23}$ transformation, still of order $\epsilon$. This is the source both of the correction to the up-quark EDM, discussed in Sect. 4.2, and of the chirality breaking $\Delta C=1$ operator proportional to $m_t$
\begin{equation}
\frac{c^{\Delta C=1} m_t}{\Lambda^2} V_{ub} V_{cb}^*g_s G^a_{\mu\nu} [(\bar{u}_L \sigma_{\mu\nu} T^a c_R)  + (\bar{u}_R \sigma_{\mu\nu} T^a c_L]  + h.c..
\end{equation}
Following \cite{Isidori:2011qw,Barbieri:2012bh},
the current limit on direct CP violation in $D\rightarrow \pi \pi, KK$ decays gives the bound
\begin{equation}
|c^{\Delta C=1}|\frac{\sin{(\delta - \phi^{\Delta C=1}})}{\sin{\delta}} \lesssim 0.2  \Big(\frac{\Lambda}{3~\text{TeV}}\Big)^2.
\end{equation}

 \begin{table}[h]
\renewcommand\arraystretch{1.6}
\hspace*{-0.05cm}
\begin{tabular}{|c|c|c|c|c|c|c|}
\hline
Observable & $\mu \rightarrow e \gamma$ & e EDM & u EDM & d EDM& $\epsilon'$& $A_{CP}^{\Delta C=1}$\\
\hline 
Coefficient &    $|c^{\mu\rightarrow e\gamma}|$                        &  $ |\text{Im}(\tilde{c}_e^{EDM})|  $  & $ |\text{Im}(\tilde{c}_u^{EDM})|$ &$ |\text{Im}(c_d^{EDM})|$       & $|c^{\Delta S=1} \text{sin}\phi|$          &  
$|c^{\Delta C=1}|$                \\
\hline
Upper bound&    $5\times 10^{-4}$    & $1.6 \times 10^{-5} $      &    $1.2\times10^{-2}$   &  $5.6 \times 10^{-3}$    & $6.5 \times 10^{-2}$
& 0.2\\
\hline
\end{tabular}
\caption{Upper bounds on the coefficients of the operators discussed in the text, normalized to $\Lambda = 3$ TeV}
\label{tab:tab1}
\end{table}

Before closing this Section we summarize in Table \ref{tab:tab1} the bounds on the coefficients of the different operators normalized to a scale $\Lambda = 3$ TeV. The bounds on $Im(c_e^{EDM})$ and $Im(c_u^{EDM})$ are $8\cdot 10^{-5}$ and $1.2\cdot 10^{-2}$ respectively. The bounds on the coefficients of the other FCNC operators are at the typical $10^{-1}\div 10^{-2}$ level\cite{Barbieri:2012uh, Barbieri:2014tja}, depending on their phases, and are the same in $U(2)^2$ or  $U(3)^2$.

\section{$U(2)^2$ in supersymmetry}

The picture that emerges from the previous Sections is that $U(2)^2$ gives rise to several new effects than the ones normally considered in MFV, with the relevant feature, as in MFV, that their flavour structure, both in the quark and, more interestingly, in the charged lepton sector, is always controlled by the CKM matrix elements.  It is this feature that allows  to make quantitative considerations\footnote{This same feature is achieved in \cite{Barbieri:1995tw}, where, however, it rests on a superpotential as in  (\ref{superpot}) without the coupling to $H_{45}$, necessary for a realistic description of fermion masses.}.

There are two good reasons that make it relevant to see the impact of the above considerations in the special case of supersymmetry. First of all is the obvious connection of supersymmetry with gauge coupling unification. Not less important, however, is the consideration of the bounds in Table \ref{tab:tab1}. Unless the various coefficients include a loop suppression factor, as in the case of supersymmetry, one can interpret them as quite strong lower bounds on the scale $\Lambda$, much above the few TeV range. In turn this appears in contrast with our original motivation of understanding the compliance with the flavour constraints of new particles with masses in the TeV range.

The model we consider is a generic supersymmetric $SU(5)$-theory with a Yukawa superpotential that gives rise to $\mathcal{L}_Y^{U(2)}$ as in eq. (\ref{gutY}) and with soft supersymmetry breaking terms generated by supergravity. In the flavour sector the entire theory is invariant under $U(2)^2$ as in eq.s (\ref{U2def}) and (\ref{U2fields}).
On this basis we shall consider the low energy theory in two different ways. We implement the general case as discussed above or we take universal  $A$-terms at least when restricted to the $1-2$ sector. 

At low energy flavour changing effects are present in the Yukawa couplings, in the   $A$-terms and in the squared masses for squarks and leptons. The Yukawa couplings $Y_{u,d,e}$, with the usual meaning of the angle $\beta$,
\begin{equation}
\mathcal{L}_Y = v \sin{\beta} \bar{u}_L Y_u u_R + v \cos{\beta} (\bar{d}_L Y_d d_R + \bar{e}_L Y_e e_R) + h.c.
\label{LY}
\end{equation}
take the form of eq. ({\ref{YuYdYe}). The $A$-terms 
\begin{equation}
\mathcal{L}_A = v \sin{\beta} \tilde{u}_L^+ A_u \tilde{u}_R + v \cos{\beta} (\tilde{d}_L^+ A_d \tilde{d}_R + \tilde{e}_L^+ A_e  \tilde{e}_R) + h.c.
\label{Aterms}
\end{equation}
have an analogous structure 
\begin{equation}\begin{split}
A_{u}=\left(
\begin{array}{c:c}
a_u \lambda_{u} & y_{t}a_{ut}\mathbf{V} \\ \hdashline
y_{t}a_{ut}\mathbf{V}^T & a_ty_{t}.
\end{array}\right)
\quad\quad
A_{d}=\left(
\begin{array}{c:c}
a_{d1}\lambda_{d} + a_{d2}\lambda_e^T & y_{b}a_{db}\mathbf{V} \\ \hdashline
0 & a_by_{b}
\end{array}\right)
\end{split}
\label{Aud-form}
\end{equation}
and
\begin{equation}
A_{e}=\left(
\begin{array}{c:c}
a_{e1}\lambda_{e} + a_{e2}\lambda_d^T & 0 \\ \hdashline
y_{\tau}a_{e\tau}\mathbf{V}^T & a_\tau y_{\tau}
\end{array}\right)
\label{A_e-form}
\end{equation}
where the various $a$-factors are mass terms of similar order of magnitude, related to the low energy scale of effective supersymmetry breaking. Finally the squared masses have two different forms. Up to negligibly small 
terms quadratic in $\lambda_{e, d, u}$, the squared masses for $\tilde{L}$ and $\tilde{d}_R$ are diagonal and degenerate in $1-2$ sector, whereas the mass terms for $\tilde{e}_R, \tilde{u}_R$ and $\tilde{Q}$ have contributions controlled by the spurion $\mathbf{V}$, i.e.
\begin{equation}
M_{\tilde{e}}^2=\left(
\begin{array}{c:c}
m_{e1}^2 \bold{1} & m_{e12}^2\mathbf{V} \\ \hdashline
m_{e12}^2\mathbf{V}^T & m_{e3}^2.
\end{array}\right)
\end{equation}
and similar for $M_{\tilde{Q}}^2, M_{\tilde{u}}^2$.

\subsection{$\mu\rightarrow e + \gamma$}

To discuss $\mu\rightarrow e \gamma$ it is convenient to go to the basis where $Y_e$ and $M_{\tilde{e}}^2, M_{\tilde{L}}^2$ are diagonal, i.e. the physical basis for the charged leptons but not for the sleptons, since $A_e$ is still non diagonal. In this basis both the bino, $\chi$, and the neutral higgsino have flavour changing interactions with the $(e_R, \tilde{e}_R)$ multiplets
\begin{equation}
\mathcal{L}_{\chi, \tilde{h}} = \sqrt{2} g^\prime \chi(\tilde{e}_R^+ U_{e}^+ e_R) + v \cos{\beta}\tilde{h} (\tilde{e}_R^+ U_{e}^+Y_e^D e_L) + h.c.,
\label{chihiggsino}
\end{equation}
where, as in previous examples, $U_e$ is as $U_{eR}$ in eq. (\ref{matru2}), except for a different unitary transformation in the $2-3$ sector, although still of order $\epsilon$. Furthermore in the $1-2$ sector $U_{eT} = U_d$ up to a small misalignment of order $m_d/m_s$. Hence the  flavour violation in these interactions is controlled by the CKM angles. Note also the non degeneracy of relative order $\epsilon^2$ between the first two generations of right-handed leptons, that will play a role in the following.

Let us now look at the $A_e$-term in eq. (\ref{Aterms}). If the $A$-terms were universal, at least in the $1-2$ sector, one would have no $\lambda_d^T$ term in (\ref{Aud-form}) and (\ref{A_e-form})
and, in the basis under consideration, it would be
\begin{equation}
A_{e} \rightarrow \left(
\begin{array}{c:c}
a_{e1}\lambda_{e}^D & 0 \\ \hdashline
0 & a_\tau y_{\tau}
\end{array}\right)  
U_e.
\label{Aele}
\end{equation}
On the contrary, in the general case in which both $\lambda_e$ and $\lambda_d^T$ are present, it is the last one that dominates. Therefore, in this case it is
\begin{equation}
A_{e} \rightarrow \hat{U}_{12}(m_d/m_s)
\left(
\begin{array}{c:c}
a_{e2}\lambda_{d}^D & 0 \\ \hdashline
0 & a_\tau y_{\tau}
\end{array}\right)  
U_{12}(m_d/m_s) U_e
\label{Aeld}
\end{equation}
where on both sides of the diagonal term there appear two unitary $1-2$ transformations of order $m_d/m_s$, representing precisely the misalignment of $\lambda_d^T$ with $\lambda_e$.

The diagrams that contribute to $\mu\rightarrow e \gamma$ are shown in Fig.s \ref{fig:fig1}. Based on eq.s (\ref{chihiggsino}) and (\ref{Aele}) (with $A$-terms universal) the only flavour changing matrix present in these interactions is $U_e$.  As such,  a GIM-like cancellation takes place, controlled by the non degeneracy of the charged sleptons, of relative order unity between the third and the first two generations and of relative order $\epsilon^2$ within the first two generations. As a consequence the $\mu\rightarrow e \gamma$ amplitude receives contributions proportional to $m_\mu U^*_e(\mu\tau)U_e(e \tau)$ (a $1-2/3$ effect) or to $m_\mu \epsilon^2 U^*_e(\mu\mu)U_e(e\mu)$ (a $1-2$ effect), both equal to $m_\mu V_{ts}^*V_{td}$ up to a factor of order unity. 
In the case of general $A_e$-term the presence of the misalignment matrices in eq. (\ref{Aeld}) inhibits the GIM-like cancellation in the diagram of Fig \ref{fig:fig1b}, which then becomes the dominant contribution to the amplitude, proportional to $m_d$ and mediated by exchanges of sleptons of the first two generations.
\begin{figure}[htbp]
\subfloat[]{\includegraphics[scale=0.55]{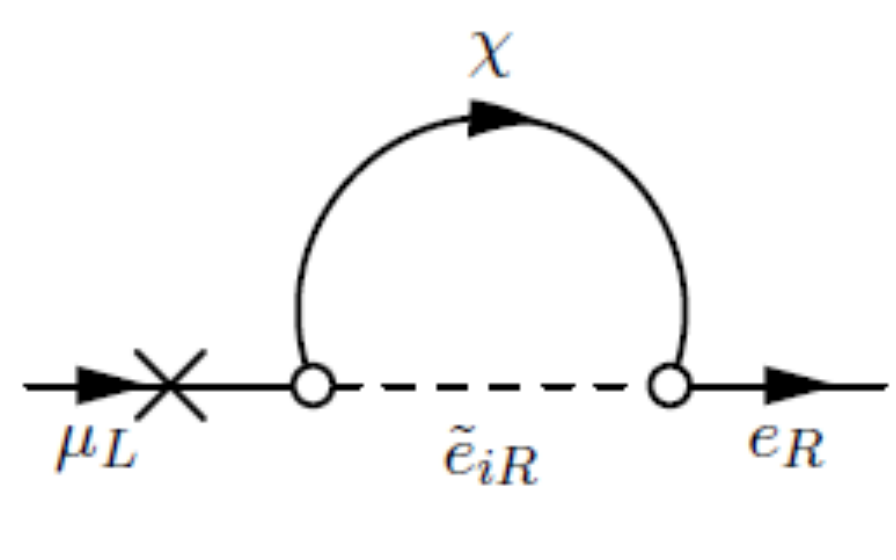}\label{fig:fig1a}}\hspace{15pt}
\subfloat[]{\includegraphics[scale=0.55]{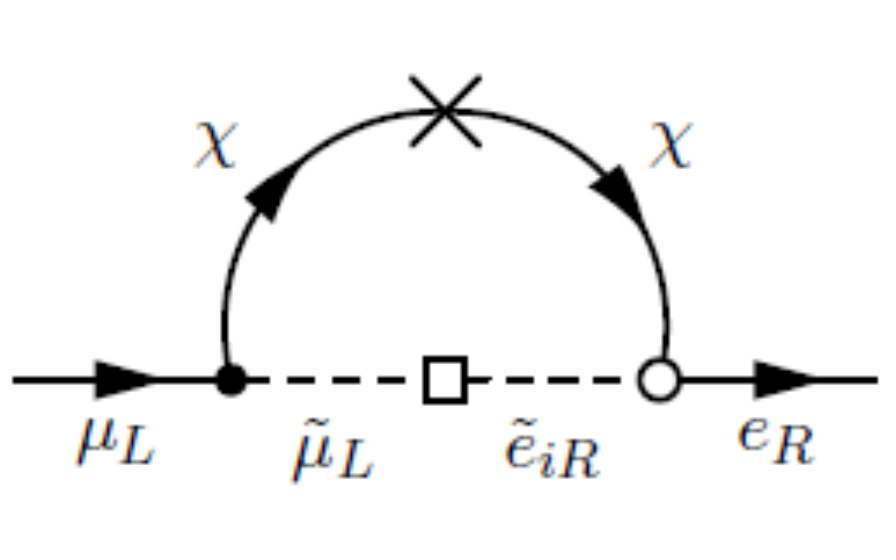}\label{fig:fig1b}}\hspace{15pt}
\subfloat[]{\includegraphics[scale=0.55]{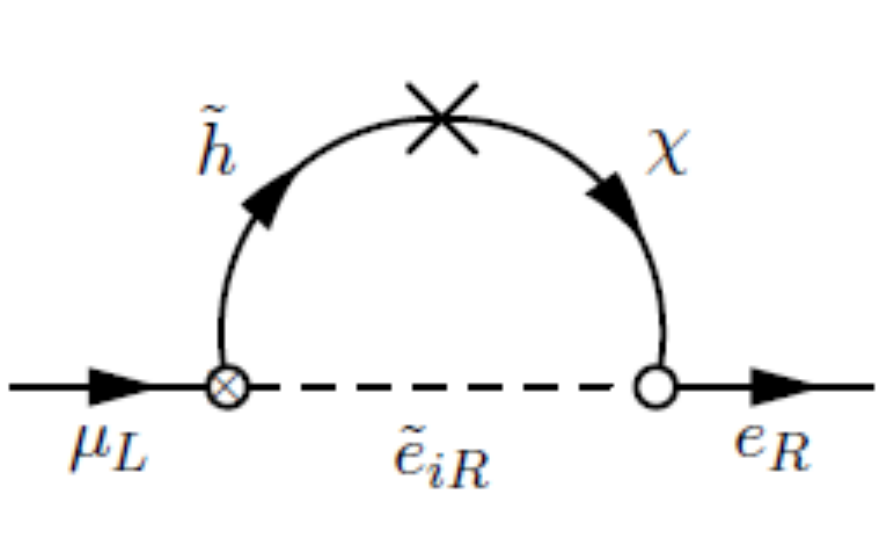}\label{fig:fig1c}}
\caption{Diagrams contributing to the $\mu\rightarrow e \gamma$ amplitude. Crosses denote a chirality flip. White circles denote flavour changing interaction vertices controlled by $U_e$. The white square in  Fig. \ref{fig:fig1b} is the $A_e$-insertion.}
\label{fig:fig1}
\end{figure}

Representative values for the size of these effects, taken incoherently, and normalized to the current limit, $BR(\mu\rightarrow e \gamma) < 5.3\cdot 10^{-13}$\cite{Adam:2013mnn}, are shown in Fig. \ref{fig:fig2},\ref{fig:fig3} both for  general and  universal $A_e$-term.  Consistently with this bound, the largest possible value of  $BR(\tau\rightarrow \mu \gamma)$ can be reached with a universal $A_e$-term, at $10^{-9}$ level. 
One should remember the order one unknown factor in front of each of these amplitudes.
\begin{figure}
 \centering
 \includegraphics{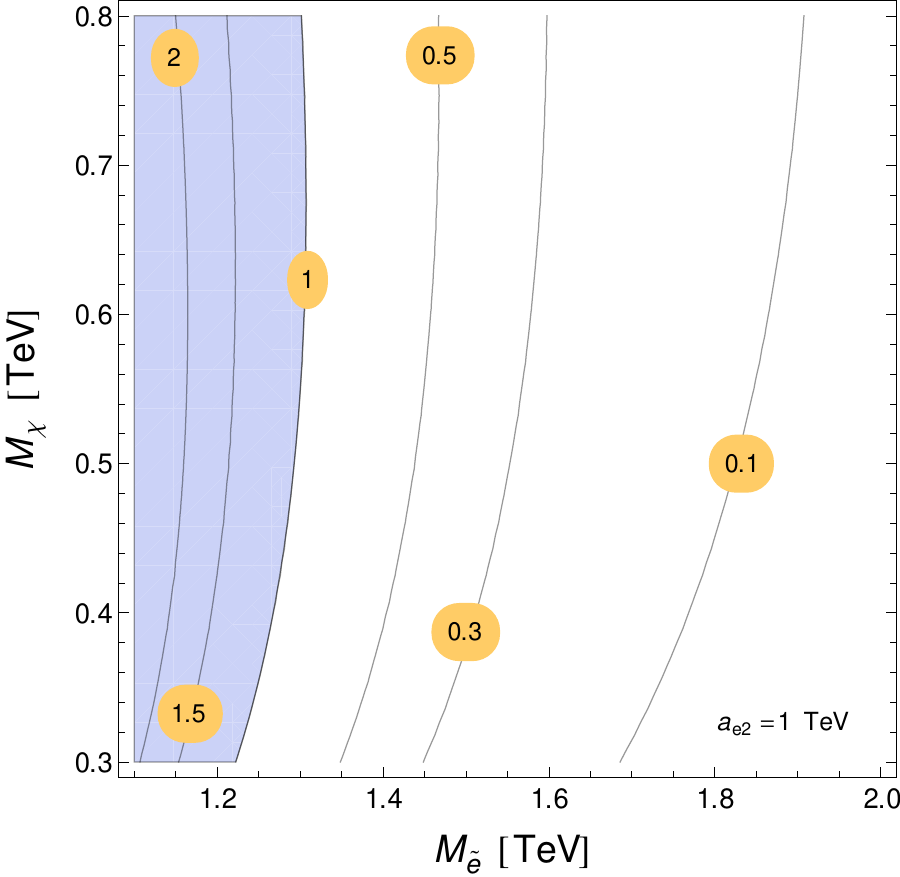}
 \caption{$BR(\mu\rightarrow e \gamma)$ normalized to the current bound, $BR < 5.3\cdot 10^{-13}$, with non universal $A$-terms for a right-handed selectron mass of the first two generations $M_{\tilde{e}}$ and neutralino mass $M_{\chi} = \mu$.}
\label{fig:fig2}
 \end{figure}
\begin{figure}
 \centering
 \includegraphics[scale=0.87]{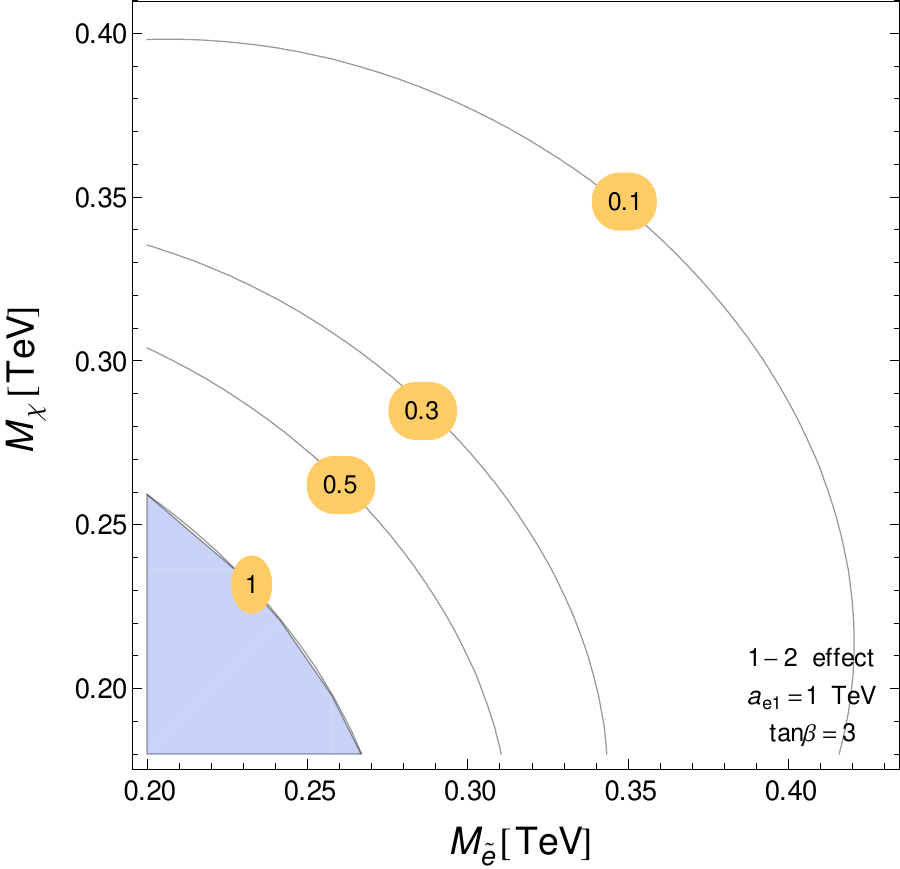}
 \hspace{8pt}
 \includegraphics[scale=0.87]{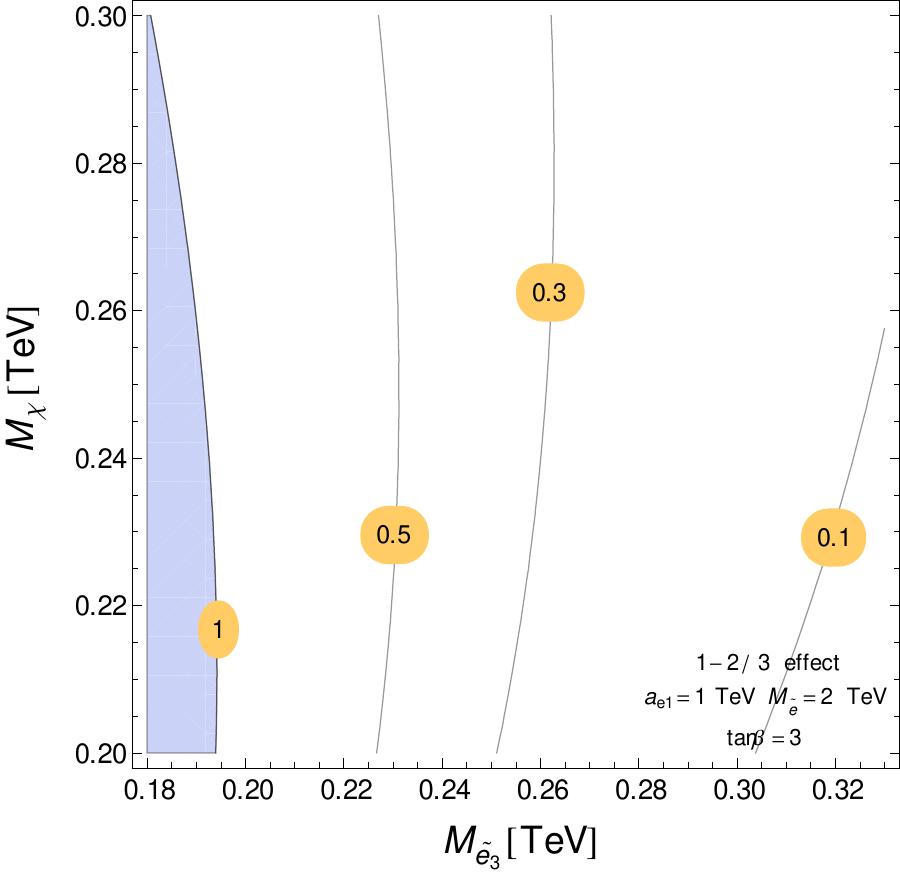}
 \caption{As in Fig. \ref{fig:fig2}  with universal $A$-terms for a right-handed selectron mass $M_{\tilde{e}}$ of the first two generations ($1-2$ effect, left) and for a  right-handed selectron mass $M_{\tilde{e}_3}$ of the first two generations ($1-2/3$ effect, right).}
\label{fig:fig3}
 \end{figure}

\subsection{Electron  EDM}

As seen in Table \ref{tab:tab1}, the electron EDM is potentially capable of providing the strongest limit, although generally dependent on  more than one unknown phase. In supersymmetry a well known effect that arises from interactions not included in eq.s (\ref{LY}) or (\ref{Aterms}), since it is not related to flavour changing phenomena, is due to a chargino-sneutrino one loop diagram. From the current experimental limit\cite{Baker:2006ts} and $|M_\chi| = |\mu|$ on obtains the bound on the mass of the sneutrino of the first generation\cite{Barbieri:2011vn} 
\begin{equation}
m_{\tilde{\nu}} > 17\hspace{2pt}\text{TeV} \cdot (\sin{\phi_\mu} \tan{\beta})^{1/2}
\end{equation}
The interactions in  eq.s (\ref{LY}) or (\ref{Aterms})  also contribute to the electron EDM by diagrams analogous to the ones in Fig. \ref{fig:fig1}, again with a distinction between universal or non universal $A_e$-term. Representative values for the size of these effects are  shown in Fig. \ref{fig:fig4},\ref{fig:fig5} for maximal values of the relevant phases and some choice of the other parameters, as indicated.
\begin{figure}
 \centering
 \includegraphics{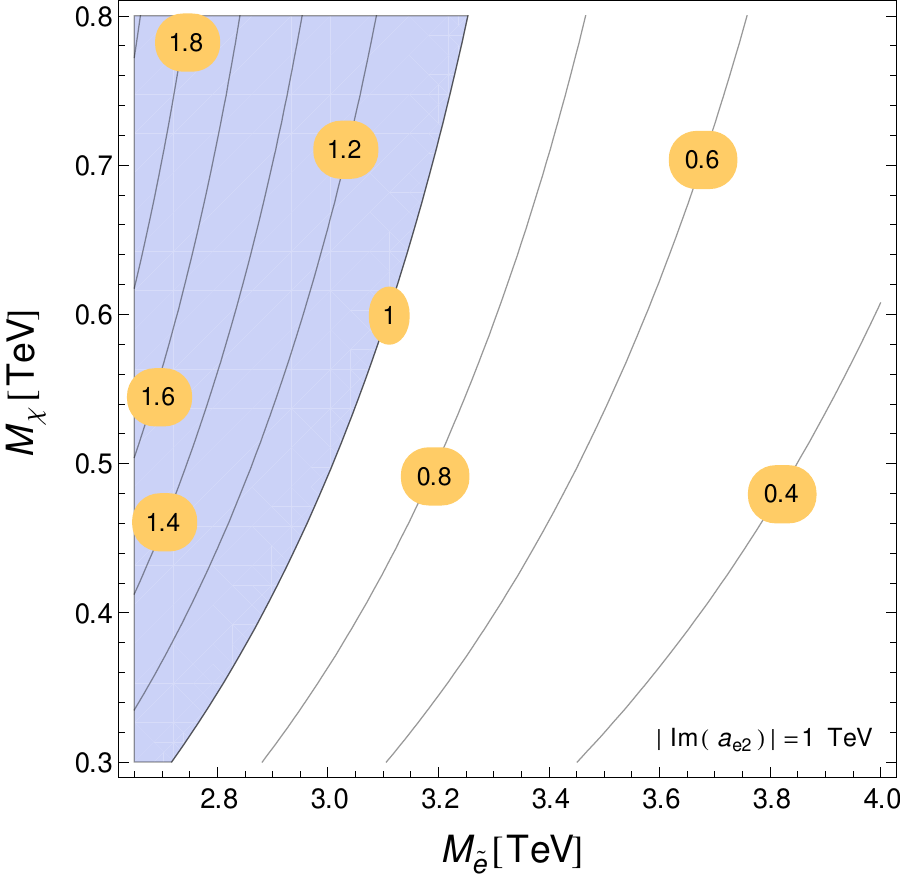}
 \caption{Electron EDM normalized to the current bound, $d_e < 8.7\cdot10^{-29} e\cdot cm$, with non universal $A$-terms and $M_{\tilde{e}}, M_{\chi}$ as in Fig. \ref{fig:fig2}.}
\label{fig:fig4}
 \end{figure}
\begin{figure}
 \centering
 \includegraphics[scale=0.87]{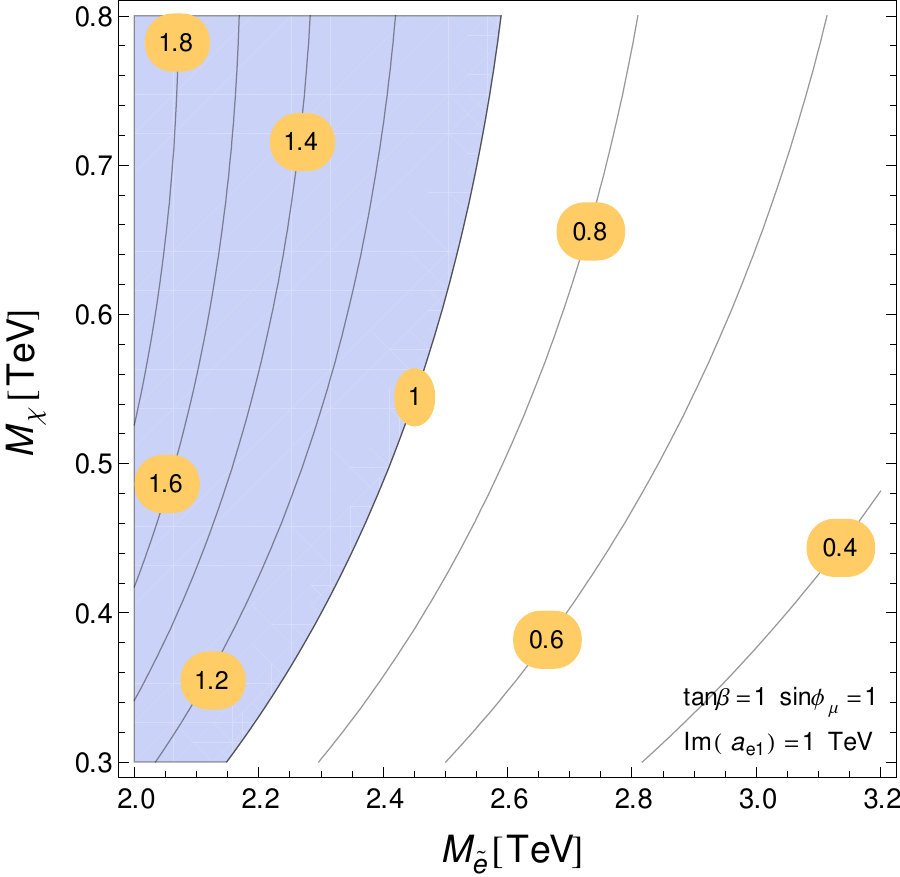}
 \hspace{8pt}
 \includegraphics[scale=0.87]{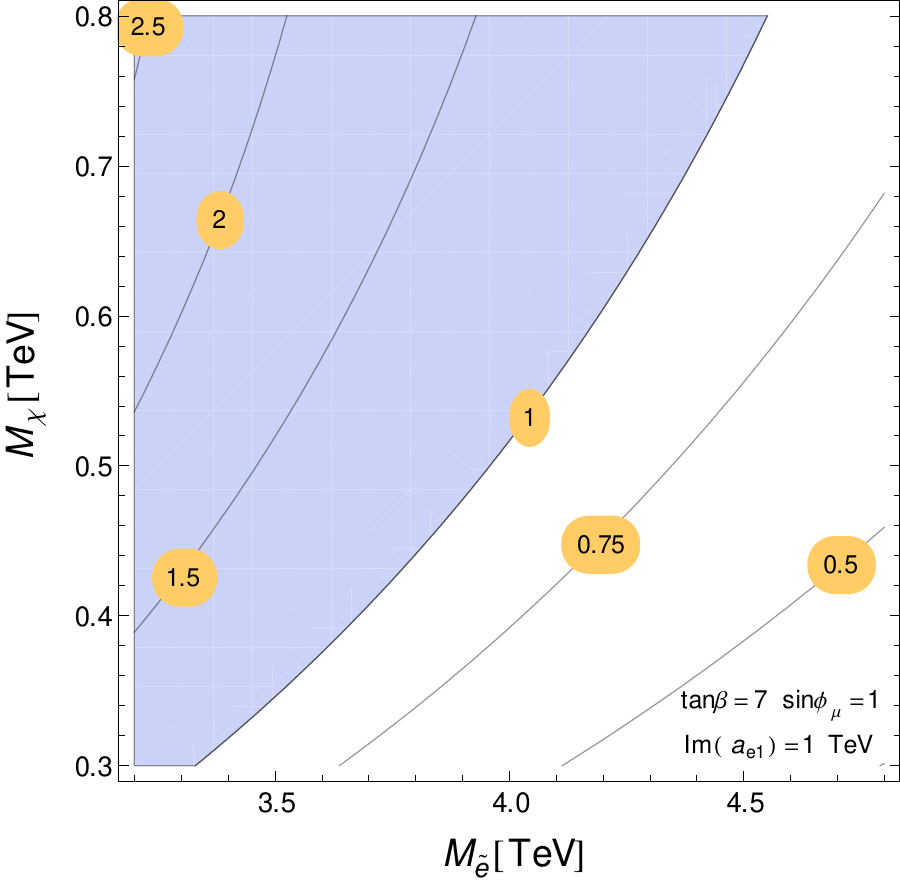}
\caption{As in Fig. \ref{fig:fig4} with universal $A$-terms for $\tan{\beta}=1$ (left) and $\tan{\beta}=7$ (right).}
 \label{fig:fig5}
 \end{figure}

\section{Summary and conclusions}

The effort to increase the precision of current flavour tests of the SM, now at the $10\div 30\%$ level, is a strongly motivated task of particle physics in itself. At the same time this effort could give indirect signals of the existence of new particle at the TeV scale, complementary to the potentiality of their direct search in high energy collisions. Although not exclusively, nevertheless a strong basis for this statement is the possibility that MFV be at work in some extension of the SM. Especially in its weak form, based on the $U(2)^3$ flavour group, phenomenological MFV can explain the absence of new signals so far, while making plausible their emergence in foreseen flavour physics experiments.

While MFV has a  predictive content in the quark sector, this is relatively less the case when one tries to extend it to the lepton  sector, due to the uncertainties related to the description of neutrino masses. To overcome this problem here we have proposed a predictive scheme based on extending MFV considerations to $SU(5)$-unification.  As far as the quark sector is concerned, weak MFV can be made consistent with  $SU(5)$-unification without introducing  new strong constraints, even though some interesting  CP-violating  effects appear both through  $\Delta S=1$ and $\Delta C=1$ chirality breaking operators. In the charged lepton sector, on the other hand, one predicts flavour violations with intensities also controlled, to a good approximation, by the CKM mixing angles. 
From a general EFT point of view Table \ref{tab:tab1} is an effective summary of our findings. As shown there, not unexpectedly, the current limits on $\mu\rightarrow e \gamma$ as on the electron EDM represent strong constraints.

Although not exclusively, supersymmetry is the obvious arena where these considerations might be of relevance. For this reason we have considered their implementation in a realistic  supersymmetric $SU(5)$-theory  with soft supersymmetry breaking terms generated by supergravity. Without any extra assumption $\mu\rightarrow e \gamma$ and the electron EDM with maximal CP-violating phases require charged sleptons of the first two generations in the $1\div 3$ TeV range, as shown in Fig.s \ref{fig:fig2} and \ref{fig:fig4} . Sleptons  of the first two generations in the few hundred GeV range can be made compatible with the flavor scheme proposed here provided the $A$-terms are universal and the CP-violating phases contributing to the electron EDM are not maximal. Third generation sleptons and neutralinos in the few hundred GeV range are in any case consistent with present bounds. This is illustrated in Fig.s \ref{fig:fig3} and \ref{fig:fig5}. The weaker bound on the mass of the third generation leptons comes from 
the fact that in any event the communication between them and the first two lepton generations is controlled by the small CKM matrix elements.

\subsubsection*{Acknowledgments}
 R.B. thanks Dario Buttazzo and Filippo Sala for discussions at the early stage of this work and Gino Isidori and David Straub for useful exchanges. F.S. thanks Diptimoy Ghosh for useful discussions. This work is supported in part by the European Programme ``Unification in the LHC Era",  contract PITN-GA-2009-237920 (UNI\-LHC) and by MIUR under the contract 2010YJ2NYW-010.
%

\end{document}